\documentclass[10pt]{article}
\usepackage[a4paper]{geometry}
\usepackage{graphicx,enumerate}
\usepackage{amsmath,amssymb}
\usepackage{authblk,cite,subcaption}
\usepackage{xfrac,authblk}

\usepackage{fancyhdr}
\usepackage[sc]{mathpazo}
\usepackage[T1]{fontenc}
\usepackage{hyperref}

\def\bbF{{\bf F}}

\def\bbx{{\bf x}}

\def\bby{{\bf y}}

\def\bM{{\bf M}}
\def\bbn{{\bf n}}

\def\bbt{{\bf t}}

\def\bV{{\bf V}}
\def\bbk{{\bf k}}

\begin{document}


\title{Platonic crystal with low-frequency locally resonant snail structures.\\
Wave trapping, transmission amplification and shielding.}

\author[1]{M. Morvaridi}
\author[2]{G. Carta}
\author[1]{M.Brun*} 
\affil[1]{Dipartimento di Ingegneria Meccanica, Chimica e dei Materiali, Universit\'{a} di Cagliari, Piazza d'Armi, I-09123 Cagliari, Italy}
\affil[2]{Department of Maritime and Mechanical Engineering, John Moores University, Liverpool, L3 3AF, U.K.}

\date{}

\maketitle
\begin{abstract}
\noindent
We propose a new type of platonic crystal. The proposed microstructured plate includes snail resonators with low-frequency resonant vibrations. The particular dynamic effect of the resonators are highlighted by a comparative analysis of dispersion properties of homogeneous and perforated plates. Analytical and numerical estimates of classes of standing waves are given and the analysis on a macrocell shows the possibility to obtain localization, wave trapping and edge waves. Applications include transmission amplification within two plates separated by a small ligament.
Finally we proposed a design procedure to suppress low frequency flexural vibration in an elongated plate implementing a by-pass system re-routing waves within the mechanical system.

\end{abstract}




\section{Introduction}
\label{Intro}

Metamaterials are microstructured media engineered to have properties that are not found in nature. 
The first models were developed in electromagnetism and optics and then extended to acoustic and elasticity \cite{Engheta2006,Capolino2009,Cai2009,Craster2012}. More recently, systems such as the Kirchhoff-Love plate equations for flexural waves, labelled as \emph{platonics} by McPhedran et al. \cite{McPhedran2009}, have been addressed. This flexural systems may show many of the typical anisotropic effects
from photonics such as ultra-refraction, negative refraction and Dirac-like cones \cite{Farhat2010,Smith2012,Torrent2013,Antonakakis2014,McPhedran2015}. 
Structured plates may also show the capability of cloaking applications \cite{Farhat2009a,Farhat2009b,Stenger2012,Misseroni2016} as a result of inhomogeneous and anisotropic constitutive properties and axial prestress \cite{Colquitt2014,Brun2014,Jones2015}.

One of the main properties of the biharmonic equation of motion governing the propagation of flexural wave is the decomposition into the Helmholtz and modified Helmholtz equations, associated respectively to the presence of propagating and evanescent waves. Such waves can be coupled via the boundary or interface contact conditions. 
In most configurations the flexural waves are led by their Helmholtz component \cite{McPhedran2009} and the homogenized equation can be of parabolic type at special frequencies \cite{Antonakakis2014,McPhedran2015}. However, short range wave scattering and Bragg resonance can be strongly influenced by the evanescent waves.  

Periodic structures play a major role in this field \cite{Brillouin1946}, since they create band gaps. These are frequency ranges where waves cannot propagate through the periodic system leading to possible application as acoustic and mechanical wave filters, vibration isolators, seismic shields.
Partial band gap can lead to anisotropic wave response that can be used to obtain focusing and localization \cite{Brun2010,Bigoni2013,PiccolroazMovchanCabras2017b} as well as polarization properties \cite{Lai2011,Ma2016}.

Two physical mechanisms can open band gaps: Bragg scattering and local resonance \cite{Mead1996,Hussein2014}.
Bragg scattering is associated to the generation of band gaps at wavelengths of the same order of the unit cell around frequencies governed by the Bragg condition $a=n (\lambda/2)$, ($n=1,2,3,\cdots$), where $a$ is the lattice constant of the periodic system and $\lambda$ the wavelength \cite{Movchan2009}. 
Local resonances are associated to internal resonances due to the microstructures, they can be obtained from array of resonators as suggested in the seminal work \cite{Liu2000}. Local resonances open tiny band gaps that can be at low frequencies \cite{Xiao2012,Xiao2013,Haslinger2017} and inertial amplification mechanism that can widen stop band intervals have been proposed in \cite{Acar2013,Frandsen2016}.

\begin{figure}[ht!]
\centering
{\includegraphics[width=6.2cm,angle=0]{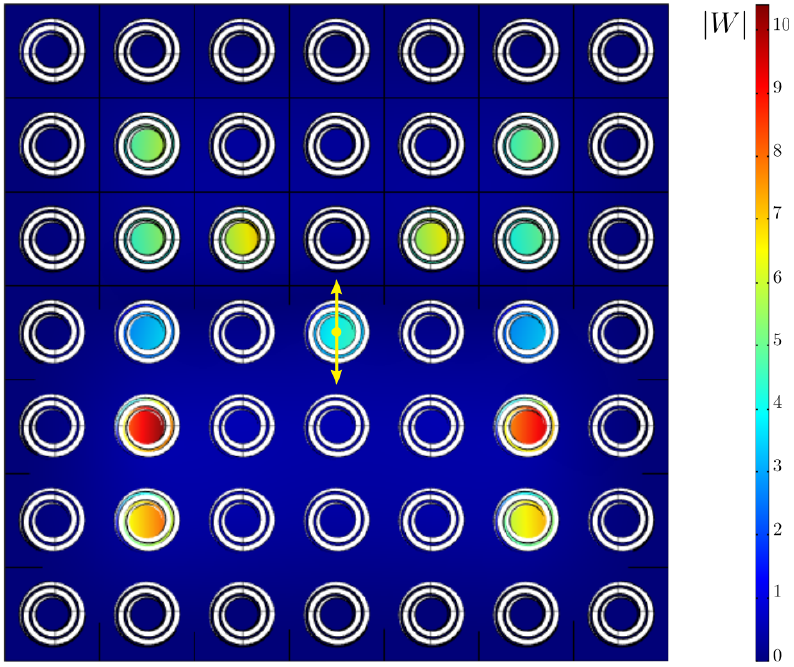}}~~~~~~
{\includegraphics[width=6.2cm,angle=0]{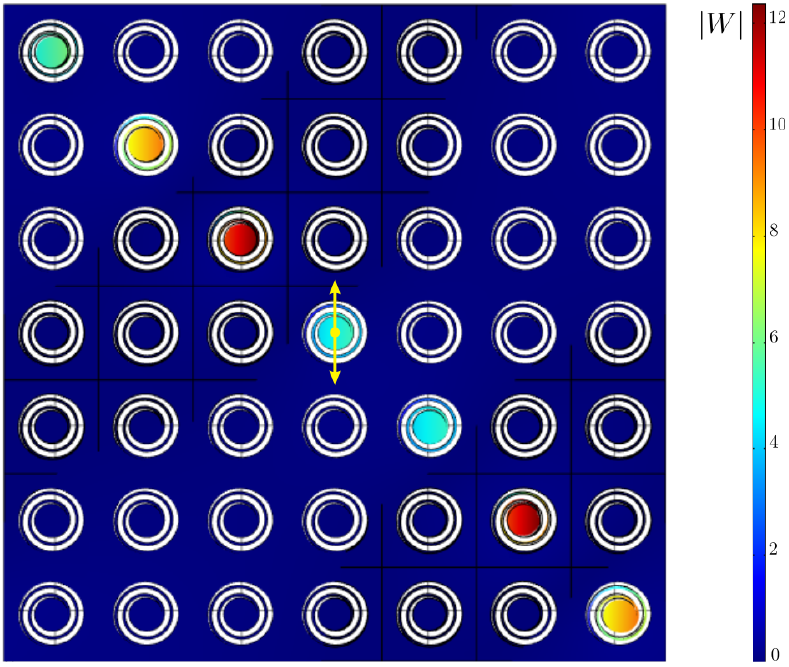}}
\vspace{5 mm}
\centerline{(a) ~~~~~~~~~~~~~~~~~~~~~~~~~~~~~~~~~~~~~~~~~~~~~~ (b) ~~~~~~}
\centering{\includegraphics[width=6.2cm,angle=0]{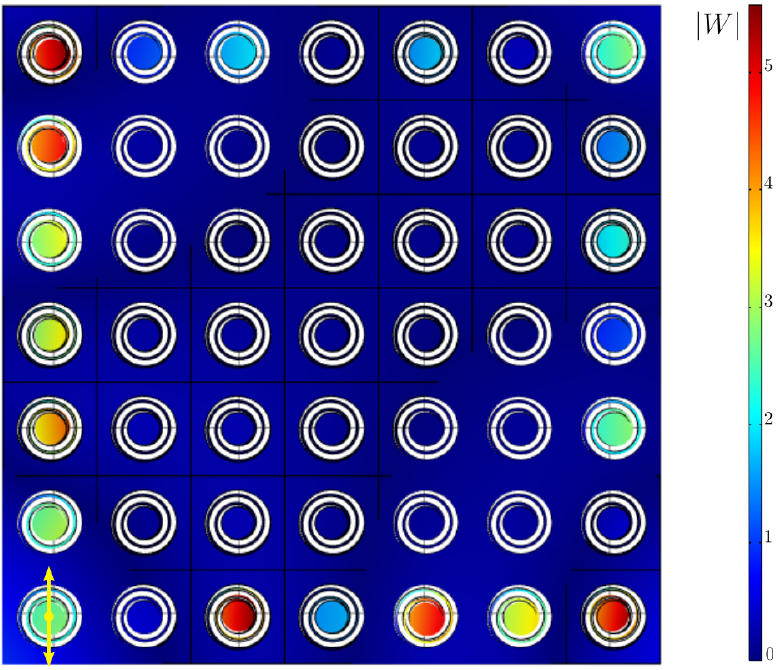}}
\centerline{(c)~~~~~~}  
 \caption{\footnotesize `Trapped' and `edge' modes in the microstructured plate. Trapped modes showing high amplitude vibrations concentrated on a letter `M' (a) or on a line (b) in a macrocell of the periodic system. The systems are excited by a transverse time-harmonic unit force applied in the center of the central inclusion. (c) Edge mode in a finite size plate composed of of $7\times 7$ unit cells. The plate is excited by a transverse time-harmonic unit force applied in the center of the bottom left inclusion.}
   \label{Fig00}
\end{figure}

\section{The platonic system of snail resonators}

We consider flexural vibrations in Kirchhoff plates. In the time-harmonic regime the transverse displacement $W({\bbx})$ satisfy the fourth-order biharmonic equation
\begin{equation}
\nabla^4 W({\bbx})-\beta^4W({\bbx})=0, \quad  \beta^4=\frac{\rho h}{D}\omega^2.
\label{eqn001}
\end{equation}
In Eq. (\ref{eqn001}) ${\bbx}=(x_1,\, x_2)^T$ is the position vector, $\omega$ is the radian frequency, $\rho$ the mass density and $D$ the flexural stiffness. Also, $D=E h^3/(12(1-\nu^2))$, with $E$ the Young's modulus, $\nu$ the Poisson's ratio and $h$ the plate thickness. We consider a steel plate, with $\rho=7800$ kg/m$^3$, $E=2\times 10^{5}$ MPa, $\nu=0.3$ and $h=1$ mm. The shear modulus is $\mu$

Rotation is the gradient vector ${\bf \phi}({\bbx})=\nabla W(\bbx)$, while the static quantities are the bending moment symmetric tensor $\bM$ and the shear force vector ${\bf Q}=\nabla\cdot\bM$.
We report the moment-curvature relations:
\begin{eqnarray}
\nonumber
M_{11} = -D \left( W_{,11} + \nu W_{,22}\right),\\
\nonumber
M_{22} = -D \left( W_{,22} + \nu W_{,11}\right),\\
M_{12} = - D (1-\nu) W_{,12}\,,
\label{eqn002}
\end{eqnarray}
where $M_{ij}$ ($i,j=1,2$) are the components of the moment tensor and the usual notation for spatial derivatives with respect to the relevant variable has been used, i.e. $W_{,i}=\partial W/\partial x_i$.
The transverse force vector $\bV$ has components $V_1=Q_1+M_{12,2}$ and $V_2=Q_2+M_{21,1}$.

\subsection{Geometry of the model}
\label{Geom}

\begin{figure}[!ht]
  \centerline{
  \includegraphics[width=13cm]{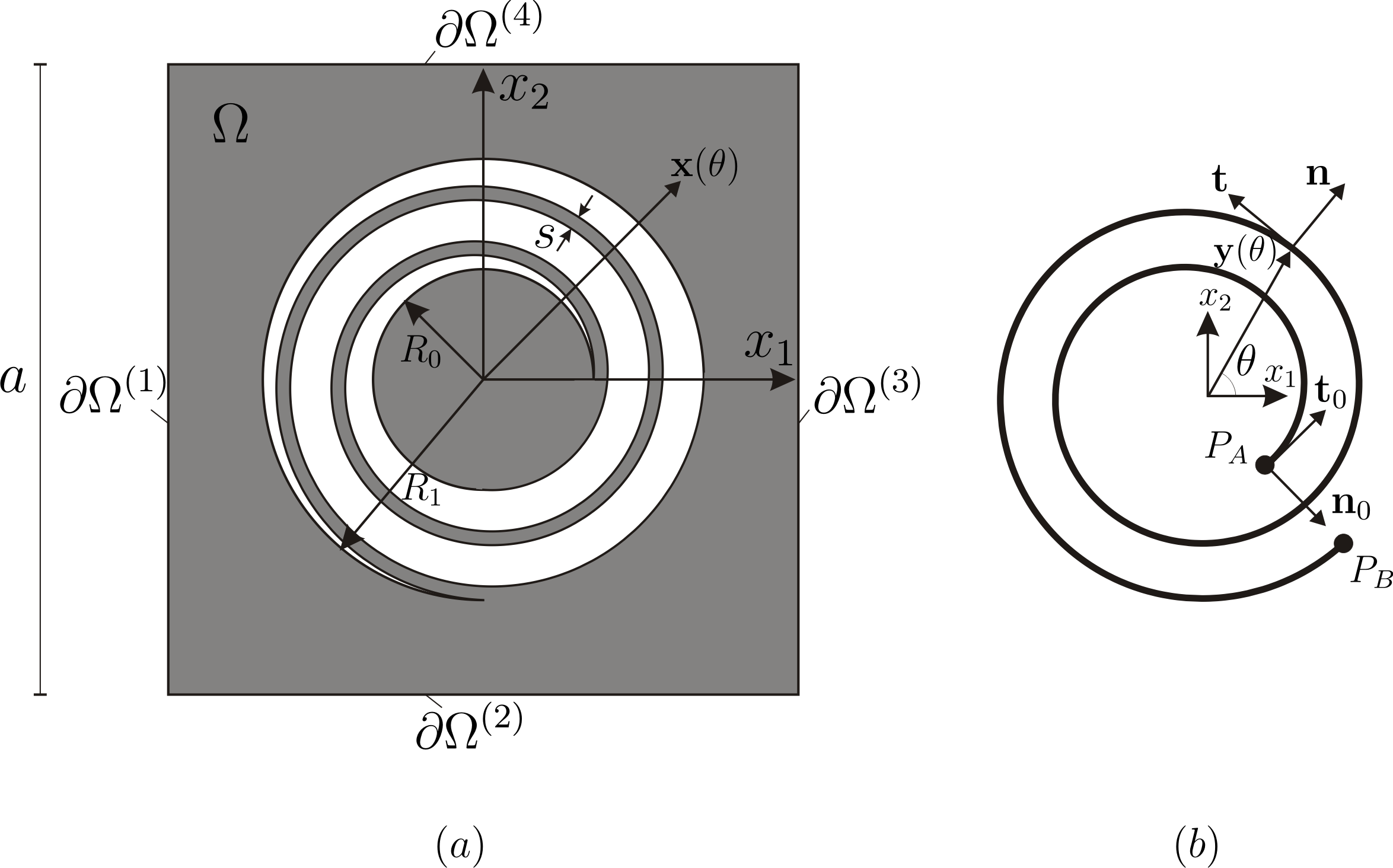}}
  \caption{\footnotesize (a) Geometry of the unit cell of the platonic crystal. (b) Geometry of the central axis of the spiral connection. Normal and tangential directions ${\bf n}(\theta)$ and ${\bf t}(\theta)$ are indicated, ${\bf n}_0={\bf n}(-\pi/4)$ and ${\bf t}_0={\bf t}(-\pi/4)$.}
  \label{Fig01}
\end{figure}

The unit cell of the square periodic system is shown in Figure \ref{Fig01}. In the unit cell of dimension $a=1$ m, a central hole of radius $R_1=0.35$ m is introduced with a circular inclusion of radius $R_0=0.175$ m, the inclusion is connected to the external structure by a slender spiral of thickness $s=21.875$ mm. 

The radial position of the central axis of the spiral is indicated in Figure \ref{Fig01}b and given by:
\begin{eqnarray}
\label{eqn003}
y(\theta)=|{\bf y}(\theta)| = R_0+(R_1-R_0)\frac{\theta+\pi/4}{4\pi},  \quad 
\theta\in\left[-\frac{\pi}{4},4\pi-\frac{\pi}{4}\right].
\end{eqnarray}

The curve can be given in the parametric form
\begin{equation}
C(\theta):={\bf y}(\theta)=[y_1(\theta),y_2(\theta)]=[y(\theta)\cos(\theta),y(\theta)\sin(\theta)],
\label{eqn003a}
\end{equation}
and normal and tangent vector to the curve are:
\begin{equation}
{\bf n}(\theta)=\frac{1}{|C'(\theta)|}
\left[
\begin{array}{c}
y_2'(\theta)\\ 
-y_1'(\theta)
\end{array}
\right], \qquad
{\bf t}(\theta)=\frac{1}{|C'(\theta)|}
\left[
\begin{array}{c}
y_1'(\theta)\\ 
y_2'(\theta)
\end{array}
\right],
\label{eqn003c}
\end{equation}
with 
\begin{equation}
|C'(\theta)|=\sqrt{y'(\theta)^2\!+\!y(\theta)^2}=\sqrt{\bigg(\frac{R_1\!-\! R_0}{4\pi}\bigg)^2\!+\!\left[R_0\!+\! (R_1\!-\! R_0)\frac{\theta+\pi/4}{4\pi}\right]^2}.
\label{eqn003d}
\end{equation}

The spiral length is
\begin{eqnarray}
\nonumber
L=\int_0^{4\pi}\!\!\!\sqrt{x'(\theta-\pi/4)^2\!+\!x(\theta-\pi/4)^2}\,d \theta = \\
\nonumber
\frac{R_1\sqrt{(R_1-R_0)^2+(4\pi R_1)^2}-R_0\sqrt{(R_1-R_0)^2+(4\pi R_0)^2}}{2(R_1-R_0)}+ \\
\frac{R_1-R_0}{8\pi} \log \left[ \frac{4\pi R_1+\sqrt{(R_1-R_0)^2+(4\pi R_1)^2}}{4\pi R_0+\sqrt{(R_1-R_0)^2+(4\pi R_0)^2}}\right],
\label{eqn003e}
\end{eqnarray}
which takes the value $L=3.303$ m.

\section{Dispersion diagram of the model}

\begin{figure}[!ht]
\centerline{
\includegraphics[width=13cm]{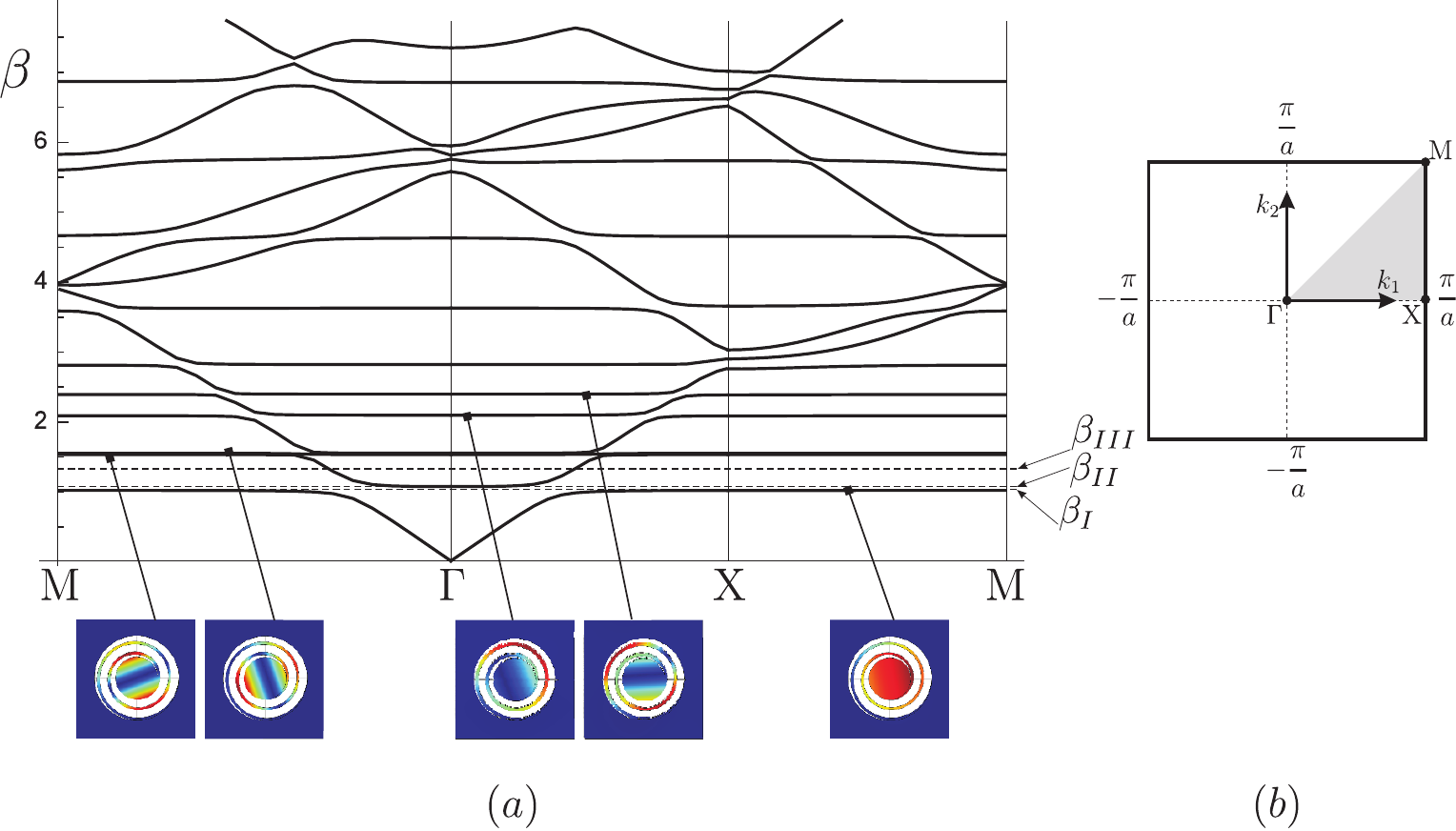}}
\caption{\footnotesize Dispersion diagram of the platonic crystal with snail resonators. The frequency parameter $\beta$ is given as a function of the wave vector $\bbk$ along the boundary of the irreducible Brillouin zone. First five localized modes are given in the bottom part. Colors from blue to red correspond to increasing amplitude of transverse displacement. (b) Sketch of the Brillouin zone in ${\bf k}$ space, with the irreducible Brillouin zone shaded. The symmetry points $\Gamma$, X and M are shown, corresponding to ${\bf k} = (0, 0)$, ${\bf k} = (\pi/a, 0)$ and ${\bf k} = (\pi/a,\pi/a)$, respectively.}
\label{Fig02}
\end{figure}

The band structure of the flexural system is presented in Figure \ref{Fig02}. The dispersion diagram has been computed performing an eigenfrequency analysis in the Finite Element package \emph{Comsol Multiphysics}\textsuperscript{\textregistered} (version 5.2) applying the following Floquet-Bloch conditions on the boundary (shown in Figure \ref{Fig01}a):
\begin{eqnarray}
\label{eqn004}
\nonumber
{W}|_{\partial \Omega^{(3)}}=e^{ik_1a}W|_{\partial \Omega^{(1)}}, \, {\phi_1} |_{\partial \Omega^{(3)}}=e^{ik_1a}{\phi_1}|_{\partial \Omega^{(1)}},\\
\nonumber
{M_{11}}|_{\partial \Omega^{(3)}}=e^{ik_1a}M_{11}|_{\partial \Omega^{(1)}}, \, {V_1} |_{\partial \Omega^{(3)}}=e^{ik_1a}{V_1}|_{\partial \Omega^{(1)}}, \\
\nonumber
{W}|_{\partial \Omega^{(4)}}=e^{ik_2a}W|_{\partial \Omega^{(2)}}, \, {\phi_2} |_{\partial \Omega^{(4)}}=e^{ik_2a}{\phi_2}|_{\partial \Omega^{(2)}},\\
{M_{22}}|_{\partial \Omega^{(4)}}=e^{ik_2a}M_{22}|_{\partial \Omega^{(2)}}, \, {V_2} |_{\partial \Omega^{(4)}}=e^{ik_2a}{V_2}|_{\partial \Omega^{(2)}},
\end{eqnarray}
where ${\bf k}=(k_1,k_2)^T$ is the wave number vector.

The dispersion diagram is given following the usual path on the boundary of the irreducible Brillouin zone sketched in Figure \ref{Fig02}b.

\begin{figure}[!ht]
\centerline{
\includegraphics[width=13cm]{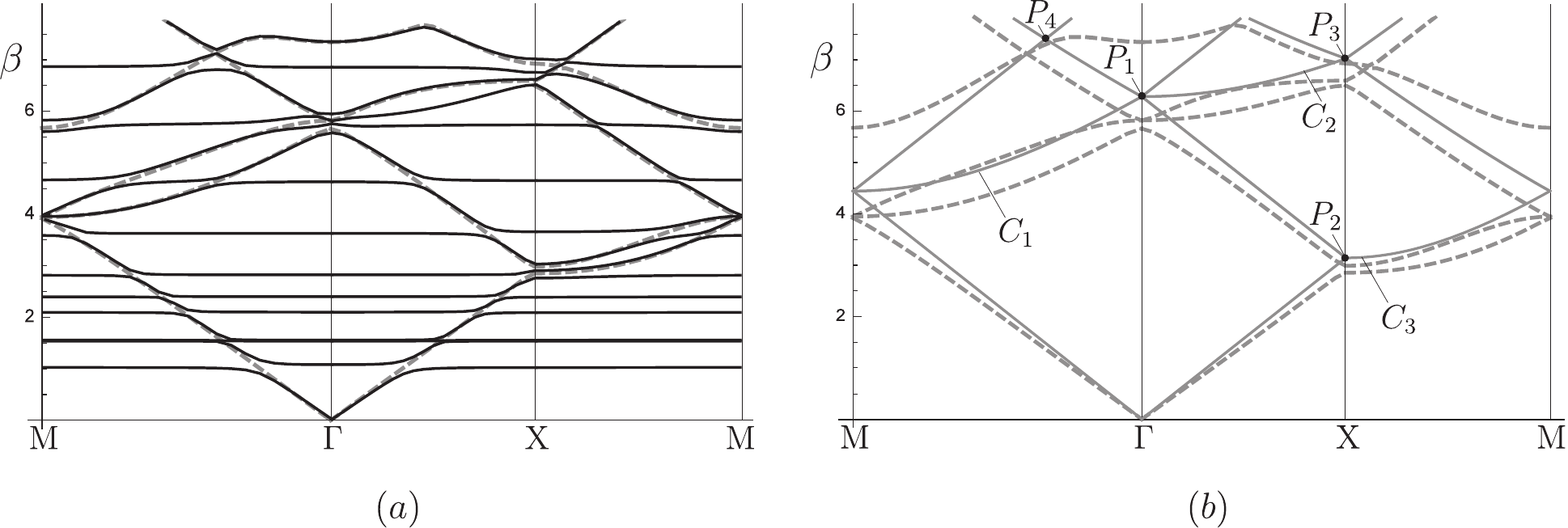}}
\caption{\footnotesize Comparison of dispersion diagrams. (a) Periodic system with snail resonators as in Figure \ref{Fig01} (black continuous lines) vs periodically perforated plate (grey dashed lines). (b) Periodically perforated plate (grey dashed lines) vs homogeneous plate (continuous grey lines).}
\label{Fig03}
\end{figure}

A comparison with the dispersion diagram of a periodic perforated plate with only the internal hole of radius $R_1=0.35$ m is given in Figure \ref{Fig03}a, while a comparison of dispersion diagrams of a perforated and homogeneous plate is given in part (b) of the same Figure. The band structure of perforated plates with free and clamped boundary conditions have been extensively studied by applying the multipole expansion method in \cite{Movchan2007, McPhedran2009,Poulton2010a, Movchan2009, Smith2014}.

The comparative analysis in Figure \ref{Fig03}b  shows that the introduction of circular perforations in a homogeneous plate induces a softening of the behavior as expected on physical ground. Additionally, it is evident the splitting between dispersion curves (see, for examples curves $C_1$, $C_2$ and $C_3$), with the formation of partial band gaps associated with wave propagation along specific directions. Also, the band gap opening at intersection points $P_1$ at $\Gamma$ and $P_2$ and $P_3$ at X is noted, while intersection points $P_3$ and $P_4$ change their position (and the frequency) in the Brillouin space.

The presence of internal resonators (Figures \ref{Fig02} and \ref{Fig03}a) gives rise to several localized mode associated to nearly horizontal dispersion curves whereas dispersion curves of the corresponding perforated plate remain practically unperturbed. First, we note that the lower localized modes appear at extremely low frequencies in correspondence of the acoustic modes of the perforated plate. Second we stress the possibility to have a large number of such flat curves spread on a large frequency interval. This model generalizes the effect on a 
single frequency previously shown in \cite{Bigoni2013}. It can be also considered as an alternative to the challenging problem of opening large stop bands at low frequency for vibration and acoustic isolation \cite{Baravelli2013, Chronopoulos2017, Frandsen2016}.

\section{Asympotic estimates for a single resonator}
\label{SemiAnaly}

The frequencies of the first internal resonance modes are estimated analytically. We asses a class of standing waves in a periodic system containing inclusions with the structured spiral coating. We assume that the inclusion at the center is taken as rigid and the the connecting spiral is an elastic massless beam.  The vibration modes in this simplified mechanical model are obtained via the introduction of the transverse displacement $W_m$ and rotations $\phi_n$ and $\phi_t$ of the rigid inclusions. The rotations $\phi_n$ and $\phi_t$ are taken around two orthogonal directions, respectively parallel to the normal ${\bf n}_0$ and tangent ${\bf t}_0$ at the intersection point $P_A$ between inclusion and spiral connection (see Figure \ref{Fig01}b).
In the asymptotic approximation of the first three eigenfrequencies, it is feasible to assume the circular contour at $x=R_1$ as rigid, so that the spiral is clamped at $P_B$. 

The kinetic energy of this mechanical system is
\begin{equation}
{\cal K}(t) = \frac{1}{2} \rho \pi \, h R_0^2\, \dot{W}^2_m+ \frac{1}{2} \rho \pi \frac{h R_0^4}{4} \left(\dot{\phi}_n^2+\dot{\phi}_t^2\right),  
\label{eqn101}
\end{equation}
whereas the potential energy, referred to the assumed massless elastic spiral is
\begin{equation}
{\cal P}(t) = \frac{1}{2} \kappa_1\, {W}^2_m+ \frac{1}{2} \kappa_2 \phi_n^2+ \frac{1}{2} \kappa_3 \phi_t^2,  
\label{eqn102}
\end{equation}
where $\kappa_1$, $\kappa_2$ and $\kappa_3$ are the elastic constrained guaranteed by the spiral to be determined in the following.

The Euler-Lagrange equations, imposing to the balance of linear and angular momenta, have the simple form
\begin{eqnarray}
\nonumber
\rho \pi \, h R_0^2\ddot{W}_m + \kappa_1 W_m=0, \\
\nonumber
\rho \pi \frac{h R_0^4}{4}\ddot{\phi}_n + \kappa_2 \phi_n=0, \\
\rho \pi \frac{h R_0^4}{4}\ddot{\phi}_t + \kappa_3 \phi_t=0,
\label{eqn103}
\end{eqnarray}
which, under the time-harmonic regime assumption, give the three resonance radian frequencies
\begin{eqnarray}
\nonumber
\omega_1=\sqrt{\frac{\kappa_1}{\rho \pi \, h R_0^2}}, \\
\nonumber
\omega_2=2\sqrt{\frac{\kappa_2}{\rho \pi \,h R_0^4}}, \\
\omega_3=2\sqrt{\frac{\kappa_3}{\rho \pi \,h R_0^4}}.
\label{eqn104}
\end{eqnarray}

For the determination of the stiffnesses $\kappa_{i}$ $(i=1,2,3)$ we make use of the Virtual Work Principle by considering the effect of flexural, torsional and shear deformation of the spiral considered as a curved beam clamped at $P_B$ (see Figure \ref{Fig01}b). 

We solve the static problem applying different concentrated loads at $P_A$.
For the determination of $\kappa_1$ we apply a force ${\bf F}=-F {\bf e}_3$, with $F=1$ N. The resulting moment $\bM^{(1)}(\theta)=(\bby(-\pi/4)-\bby(\theta))\times \bbF$, with ${\bf y}(\theta)$ given in Eq. (\ref{eqn003a}), is decomposed into bending and torsional components, $M_b^{(1)}(\theta)=-\bM^{(1)}(\theta)\cdot \bbn(\theta)$ and $M_t^{(1)}(\theta)=-\bM^{(1)}(\theta)\cdot \bbt(\theta)$, respectively. The corresponding transverse displacement $W^{(1)}$ in $P_A$ is computed considering as static and kinematically admissible fields the ones generated by the force ${\bf F}$, i.e.
\begin{equation}
W^{(1)} = \int_{-\pi/4}^{4\pi-\pi/4} \left[\frac{(M_b^{(1)}(\theta))^2}{EI} + \frac{(M_t^{(1)}(\theta))^2}{\mu I_p} + \frac{F^2}{\mu A^*}\right] \sqrt{x'(\theta)^2\!+\!x(\theta)^2}d\theta,  
\label{eqn105}
\end{equation}
where $I=s h^3/12=1.82292$ mm$^4$ (second moment of inertia), $I_p=s h^3/3=7.29167$ mm$^4$ (polar moment of inertia), $A^*=5/6sh=18.2292$ mm$^2$ (shear area). The corresponding transverse stiffness is 
\begin{equation}
\kappa_1 = \frac{F}{W^{(1)}} = 1.478 \,\mbox{N/m}.
\label{eqn106}
\end{equation}

For the rotational stiffnesses $\kappa_2$ and $\kappa_3$, we apply the moment $\bM^{(2)}=-M^{(2)} {\bf n}_0$ and $\bM^{(3)}=-M^{(3)} {\bf t}_0$, respectively, at the point $P_A$ and we take the normalized values $M^{(2)} =M^{(3)}=1$ Nm. Again, we derive the bending and torsional components, namely 
\begin{eqnarray}
\nonumber
M_b^{(2)}=-\bM^{(2)}\cdot \bbn=M^{(2)}\bbn_0 \cdot \bbn, \quad M_t^{(2)}=-\bM^{(2)}\cdot \bbt=M^{(2)}\bbn_0 \cdot \bbt, \\
M_b^{(3)}=-\bM^{(3)}\cdot \bbn=M^{(3)}\bbt_0 \cdot \bbn, \quad M_t^{(3)}=-\bM^{(3)}\cdot \bbt=M^{(3)}\bbt_0 \cdot \bbt,
\label{eqn107}
\end{eqnarray}
while the shear force is zero in these two cases. Then, analogously to Eq. (2) the rotations are
\begin{equation}
\phi_{n,t}=\int_{-\pi/4}^{4\pi-\pi/4} \left[\frac{(M_b^{(2,3)}(\theta))^2}{EI} + \frac{(M_t^{(2,3)}(\theta))^2}{\mu I_p}\right] \sqrt{x'(\theta)^2\!+\!x(\theta)^2}d\theta,
\label{eqn108}
\end{equation}
giving the rotational stiffnesses
\begin{equation}
\kappa_2 = \frac{M^{(2)}}{\phi_n} = 0.1337 \,\mbox{Nm},\quad \kappa_3 = \frac{M^{(3)}}{\phi_t} = 0.1339 \,\mbox{Nm}.
\label{eqn109}
\end{equation}
and the corresponding resonance radian frequencies
\begin{eqnarray}
\omega_1 = 1.403 \,\mbox{rad/s},\quad \omega_2 =  4.823\,\mbox{rad/s}, \quad \omega_3 =  4.827\,\mbox{rad/s}. 
\label{eqn110}
\end{eqnarray}

\begin{figure}[!ht]
\centerline{
\includegraphics[width=10cm]{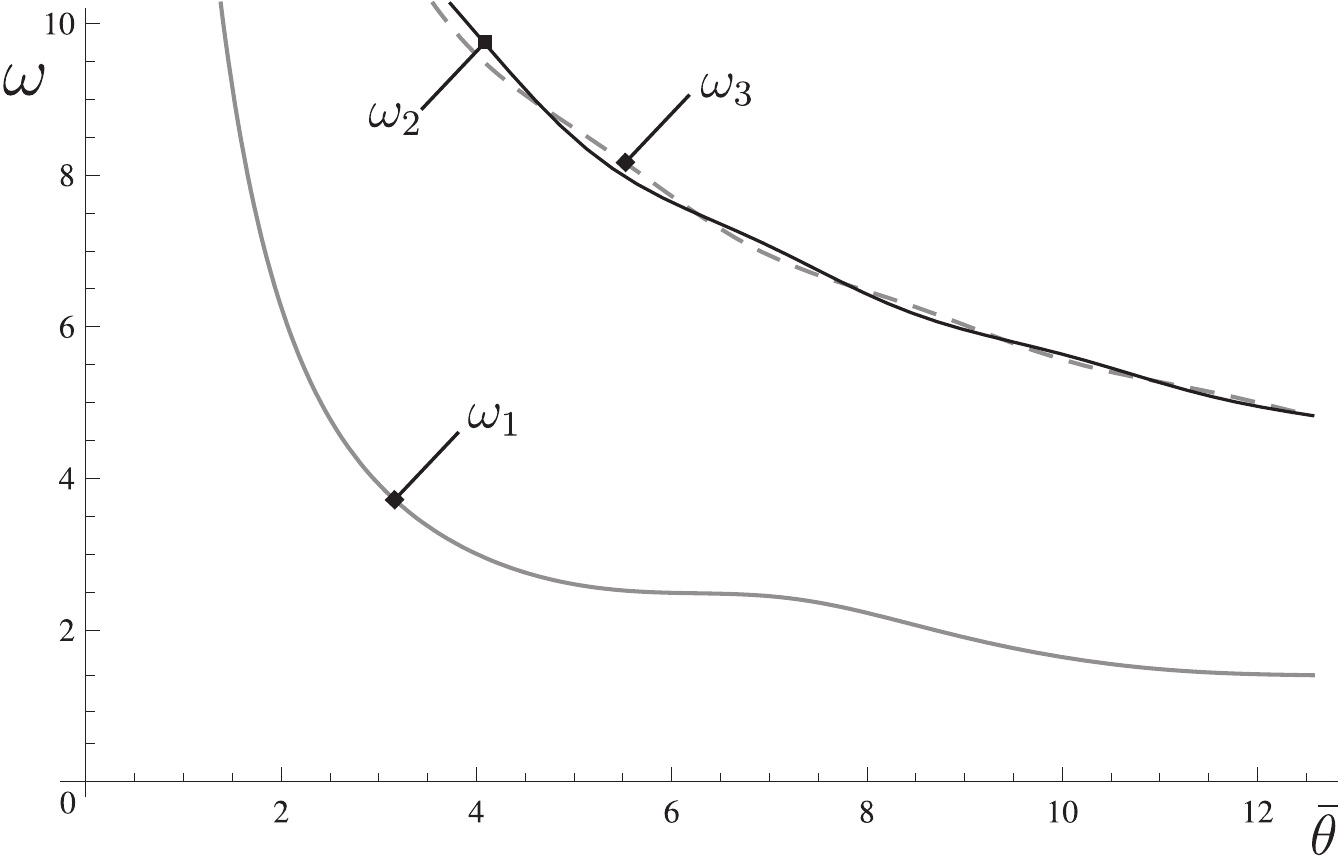}}
\caption{\footnotesize Raidan eigenfrequencies $\omega_1$, $\omega_2$, $\omega_3$ as a function of the total angular distribution $\bar{\theta}$}
\label{Fig04}
\end{figure}

We note that while shear deformation is negligible, torsional deformation has a major contribution of $\sim 80\%$ to the displacement (\ref{eqn106}) and a contribution of $\sim 40\%$ to the rotations (\ref{eqn108}). The capacity of the microstructured system to show low frequency resonance modes is strictly related to the torsional deformation of the ligament, a property that was absent in previously proposed models \cite{ Xiao2012,Colombia2015,Antonakakis2014,Haslinger2017}.

\begin{table}[!h]
  \begin{center}
    \begin{tabular}{*{10}{c}}
    \hline
    Eigenmode & (i) FEM IR & (ii) Analytical  & (iii) FEM LM \\  
    \hline
    \parbox[c]{10em}{
      \includegraphics[width=4cm]{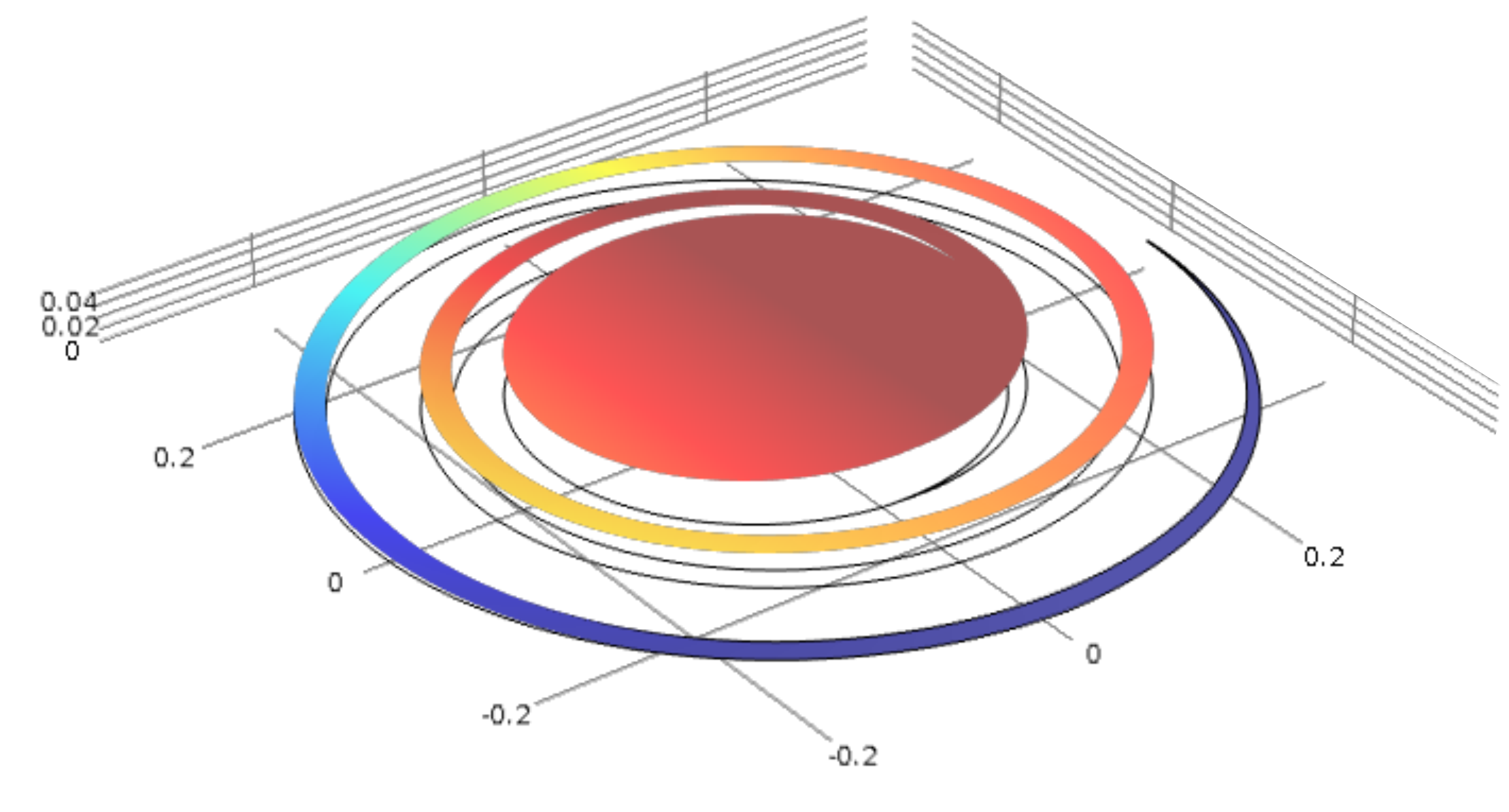}} & 1.0173 & 0.9570 & 1.0073 & \\
    \parbox[c]{10em}{
      \includegraphics[width=4cm]{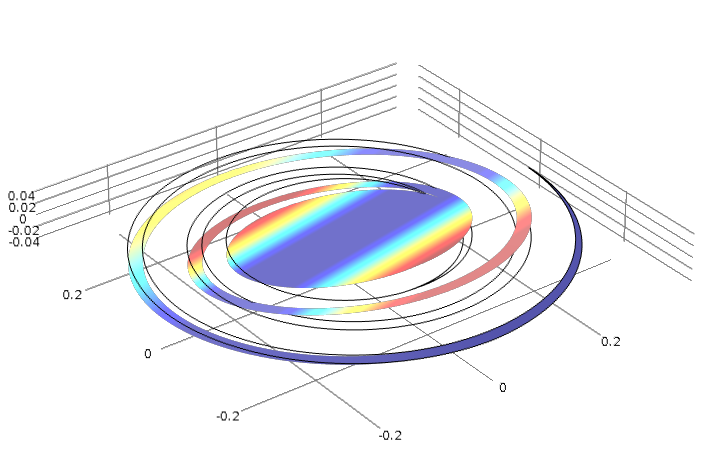}} & 1.5295 & 1.7741 & 1.5199 \\
    \parbox[c]{10em}{
      \includegraphics[width=4cm]{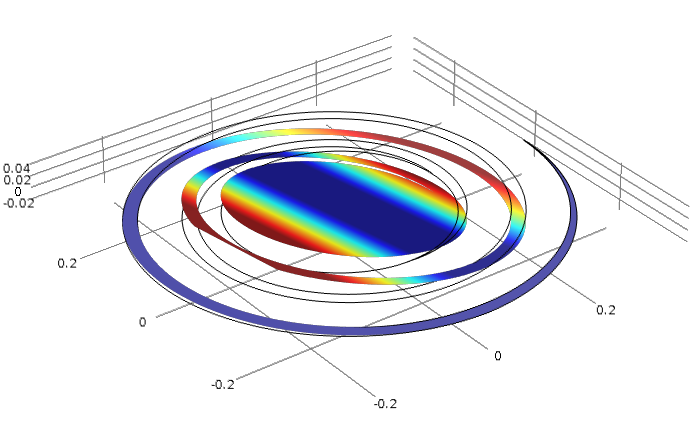}} & 1.5458 & 1.7749 & 1.5399 \\
    \parbox[c]{10em}{
      \includegraphics[width=4cm]{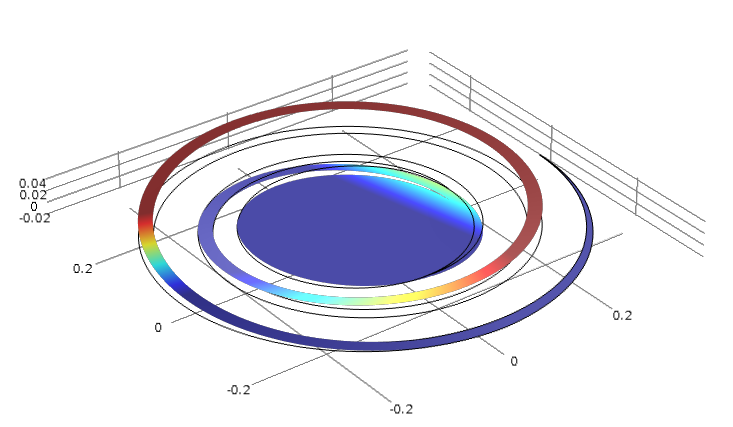}} & 2.0929 &         -  & 2.0825 \\
    \parbox[c]{10em}{
      \includegraphics[width=4cm]{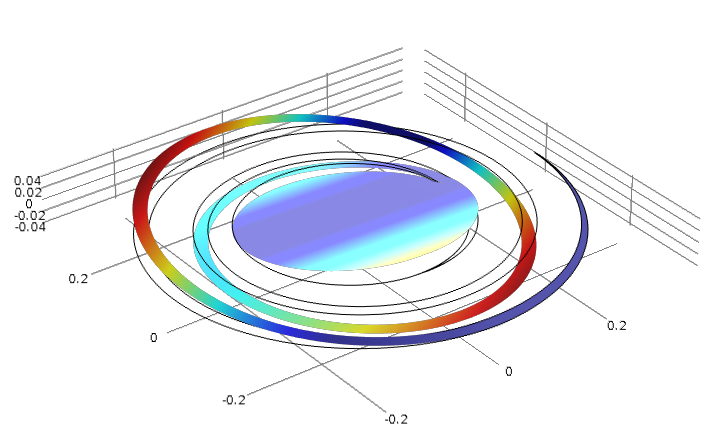}} & 2.4003 &         -  & 2.3829  &  \\
    \hline
  \hline
    \end{tabular}
  \end{center}
\caption{Eigenmodes of the isolated resonator and comparison between eigenfrequencies $\beta_i$ ($i=1,2,\ldots$) for (i) FEM solution for the isolated resonator (IR), (ii) asymptotic analytical estimates and (iii) frequencies of the localized modes (LM) in the dispersion diagram (Figure \ref{Fig01}). Colors from blue to red correspond to increasing amplitude of transverse displacement.}
\label{Table01}
\end{table}

In Figure \ref{Fig04}, we show the radian eigenfrequencies $\omega_1$, $\omega_2$ and $\omega_3$ as a function of the total angular distribution $\bar\theta$ for spirals starting at $P_A$ and having different lengths, i.e.
\begin{equation}
\label{eqn111}
y(\theta,\bar\theta)=|{\bf y}(\theta,\bar\theta)| = R_0+(R_1-R_0)\frac{\theta+\pi/4}{4\pi}, 
\theta\in\left[-\frac{\pi}{4},\bar\theta-\frac{\pi}{4}\right].
\end{equation}
Interestingly, $\omega_1$ shows some oscillations that can lead to some optimal conditions for the stiffness, in the sense that a longer spiral does not necessary give an advantage in term of achievement of a targeted low-frequency localized mode. Again, this has to be linked to the influence of torsional deformation that drastically increases the total compliance of the spiral. The torsion is not uniform and depends on the applied load and the geometry, leading to the oscillations of $\omega_i$ ($i=1,2,3$),  in Figure \ref{Fig04}. The results also evidence the large reduction rate of $\omega_1$ for $\bar\theta<\pi$.

Eigenfrequencies of an isolated continuous elastic resonator made of the spiral ligament, clamped at $y(3.75\pi)=R_1$ and the central inclusion have also been computed in \emph{Comsol Multiphysics}\textsuperscript{\textregistered}. The results confirm the analytical prediction and the results of the dispersion analysis with a first translational mode followed by two rotational modes of the central inclusion. Higher modes are associated to beam-like vibration eigenmodes of the spiral ligament \footnote{The interested reader could estimate analytically the eigenfrequency of the curved beam by implementing the approximate technique shown in \cite{Volterra1961,Volterra1961b,Volterra1961c}.}.

Eigenfrequencies $\beta_i$ ($i=1,2,\ldots$) estimated analytically are compared in Table \ref{Table01} with the eigenfrequency of the isolated resonator and the eigenfrequencies of the localized modes in the dispersion diagram of Figure \ref{Fig02}a showing a good correspondence\footnote{In the Appendix we report a second FEM analysis for a single resonator in which we show that the difference with the analytical estimated is principally due to the connection between the inclusion and the ligament.}. In particular, the agreement between two FEM analyses for single resonator and localized modes in the dispersion diagram is excellent.

\section{Numerical Results}
\label{Sect03}

\begin{figure}[ht!]
\centering
{\includegraphics[width=6.2cm,angle=0]{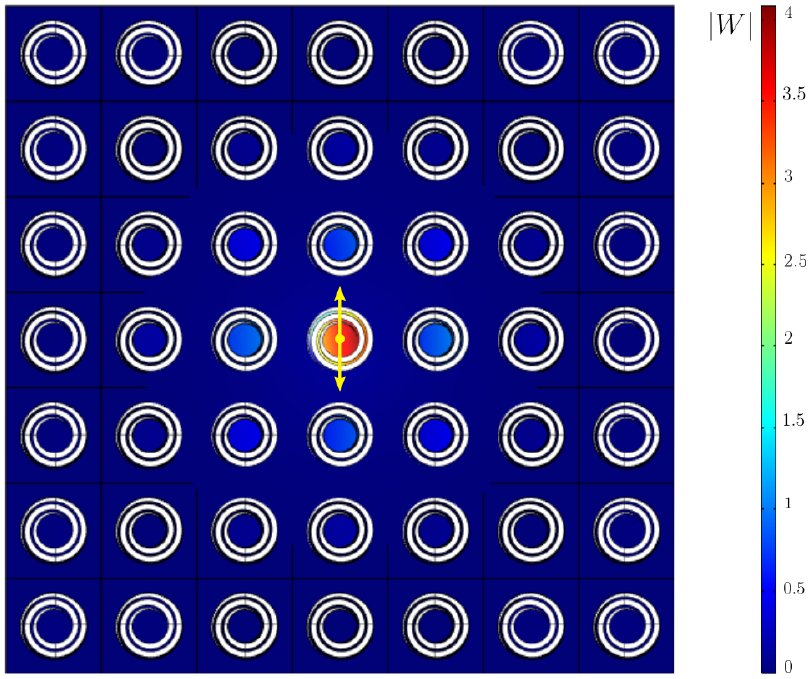}}~~~~~~
{\includegraphics[width=6.2cm,angle=0]{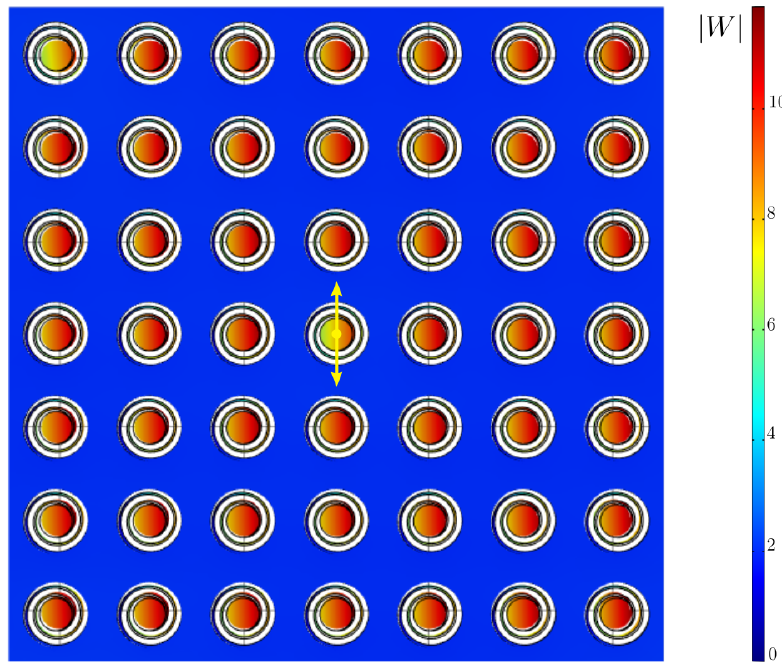}}
\vspace{5 mm}
\centerline{(a) ~~~~~~~~~~~~~~~~~~~~~~~~~~~~~~~~~~~~~~~~~~~~~~ (b) ~~~~~~}
\centering{\includegraphics[width=6.2cm,angle=0]{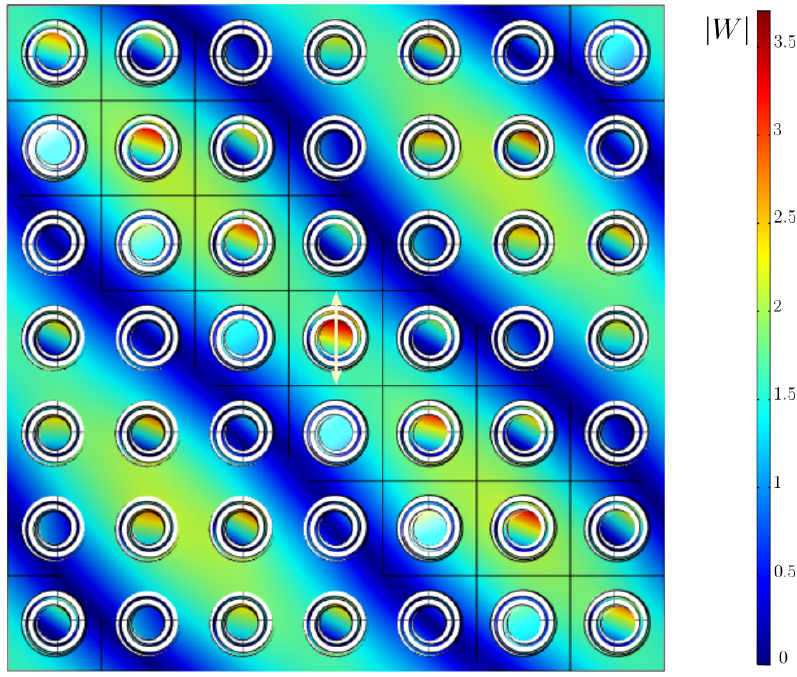}}
\centerline{(c)~~~~~~}  

  \caption{\footnotesize Periodic macrocell composed of $7\times 7$ unit cells.  Vibration modes at $\beta_I=1.0463\, \mbox{m}^{-1}$ (a), $\beta_{II}=1.0619 \,\mbox{m}^{-1}$ (b) and $\beta_{III}=1.2334 \,\mbox{m}^{-1}$. Displacement magnitude $|W|$ is shown.}
  \label{Fig07}
\end{figure}

In the following we show some numerical results, which highlight the capability of the microstructred platonic crystal to guide waves within the structure.
We start analysing the macrocell shown in Figure \ref{Fig07}, which includes 49 unit cells. The structure is subjected to a time-harmonic transverse force of amplitude $F=1$ N applied at the center of the central resonator and quasi-periodic boundary conditions.

In Figure \ref{Fig07}a we show the displacement amplitude at the frequency $\beta_I=1.0463\, \mbox{m}^{-1}$ corresponding to the first stop-band opened by the inertial resonators (see Figure \ref{Fig02}a). The strong exponential decay of the displacement amplitude is verified also for such a tiny stop band. Interestingly, the central inclusion vibrates with a translational mode whereas the next higher frequency mode, characterized by a similar exponential decay, reveals a rotational vibration of the central inclusion.

In Figure \ref{Fig07}b the frequency $\beta_I$ is slightly change to $\beta_{II}=1.0632 \,\mbox{m}^{-1}$, corresponding to the flat band around $\Gamma$ (Figure \ref{Fig02}a). In such a case, the wave propagates within the whole elastic system but only resonators vibrates with large amplitudes, while the plate undergoes a rigid displacement of negligible amplitude.  
Such an interesting behavior suggests the possibility to trap waves by properly tuning the resonance frequencies of particular sets of resonators. Such a passive system is considered in Figure \ref{Fig00}, parts (a) and (b), where a number of resonators with inclusions having mass $m=9\,\mbox{kg}$ are disposed within the macrocell. These different inclusions are disposed along the letter `M' in part (a) and along one diagonal in part (b). At the frequencies at $\beta= 0.5646\,\mbox{m}^{-1}$ and  $\beta= 0.5698\,\mbox{m}^{-1}$, respectively, they show a highly localized vibration. 

In Figure \ref{Fig00}c  we consider a different structure: a finite plate embedding $7\times 7$ unit cells is implemented with Neumann-type boundary conditions. Here, the $26$ inclusions placed in the vicinity of the external edges have different mechanical properties (mass $m=9\,\mbox{kg}$). It is evident that exciting the mechanical system with a unit vibrating force applied at the bottom-left inclusion the high amplitude vibration are localized in the vicinity of the edges.

In Figure \ref{Fig07}c the frequency is $\beta_{III}=1.233 \,\mbox{m}^{-1}$ (see Figure \ref{Fig02}). From the comparisons between the dispersion diagrams of perforated and homogeneous plates in Figure \ref{Fig03}, at $\beta_{III}$ it is evident that, at this frequency, the dispersion curves show similarities with the homogeneous case. The eigenmode reported in the Figure evidences that a plane wave propagate within the microstructure medium with low scattering, a behavior that can be linked to perfect transmission \cite{Antonakakis2014b, McPhedran2015, Neill2017}.

\begin{figure}[!ht]
\centerline{
  \includegraphics[width=13cm]{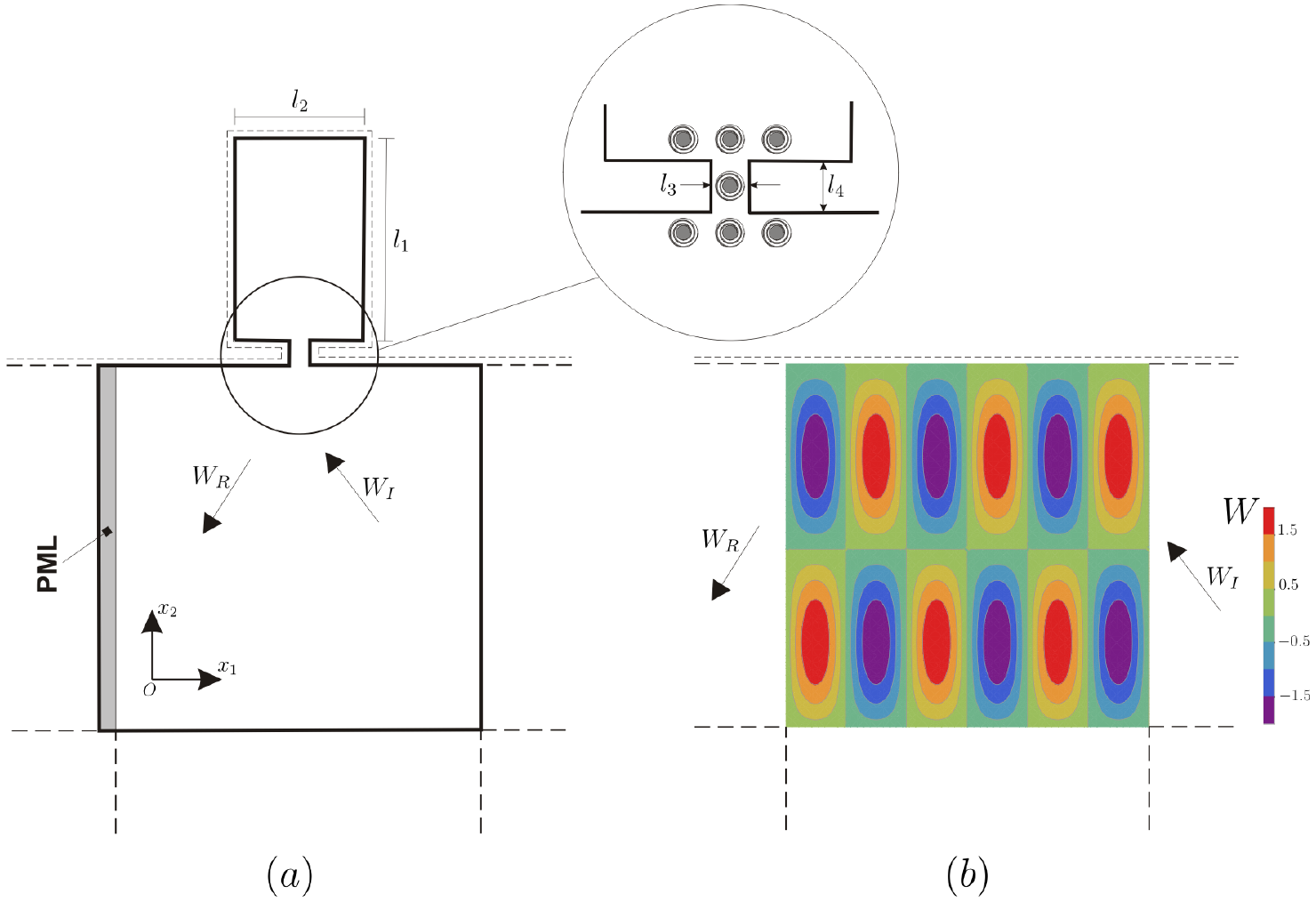}}
  \caption{\footnotesize Geometry of the system. (a) A semi infinite plate is connected to a rectangular plate with a small ligament. Seven snail resonators are placed in the vicinity of the connection. (b) Transverse displacement contours showing the solution $W=W_I+W_R$ for a semi infinite plate alone with Neumann boundary conditions. Results are given for $\beta=0.5119\,\mbox{m}^{-1}$.}
  \label{Fig08}
\end{figure}

\begin{figure}[!ht]
\centerline{
  \includegraphics[width=12.5cm]{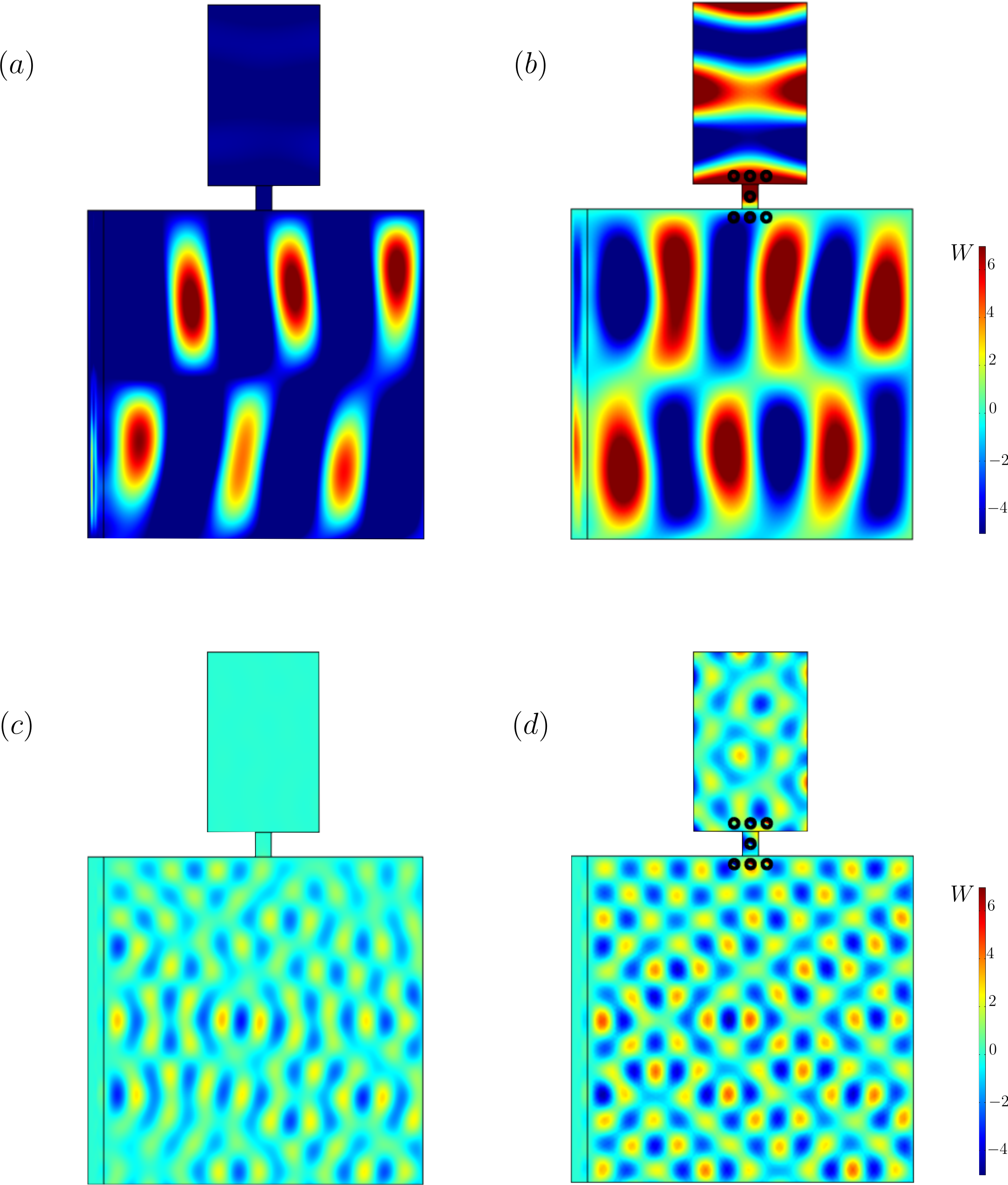}}
  \caption{\footnotesize Vibration of a semi-inifinite plate connected to a rectangular plate by means of a small ligament. Geometry of the system is given in Figure \ref{Fig08}. The structure is excited by the plane wave $W_I=e^{i \beta (x_1\cos\alpha+x_2\sin\alpha)}$, with $\alpha=11/12\pi$. The frequency is $\beta=0.5119\,\mbox{m}^{-1}$ in parts (a) and (b) and $\beta=1.4547\,\mbox{m}^{-1}$ in parts (c) and (d). (a), (c) Homogeneous plates. (b), (d) Homogeneous plate with the addition of $7$ snail resonators.}
  \label{Fig09}
\end{figure}

\subsection{Transmission amplifier}

Here we show how the microstructured plate can be implemented in order to amplify low-frequency wave transmission between two plates.
In Figure \ref{Fig08} we show the implemented geometry. A semi-infinite plate is connected to a second rectangular plate of dimensions $l_1=22\,\mbox{m}$, $l_2= 14\,\mbox{m}$ by means of a small ligament of dimensions $l_3=2\,\mbox{m}$, $l_4= 3\,\mbox{m}$. 
The mechanical and geometrical parameter of the plate are such that $\beta=2.533\sqrt{\omega} \, \mbox{m}^{-1}$ and Neumann-type boundary conditions are applied.

The system is excited by an incident plane-wave  $W_I=e^{i \beta (x_1\cos\alpha+x_2\sin\alpha)}$, with $\alpha=11/12\pi$, propagating from the semi-infinite plate. In part (b) of Figure \ref{Fig08}, we also report the analytical solution for the semi-infinite plate alone, which involves the superposition of the incident and reflected wave $W_R=e^{i \beta [x_1\cos\alpha+(\bar x_2-x_2)\sin\alpha]}$, where $x_2=\bar x_2$ defines the boundary of the plate.  

In order for the wave to penetrate in the second finite plate and produce high-scattering in the semi-infinite one, the wavelength of the incident wave must be small with respect to the width $l_3$ of the ligament. Numerical computations, not reported for brevity, indicates approximatively $(l_3 \beta \cos\alpha)>11$. Therefore it is needed to reach sufficiently high frequency in order to enhance transmission of waves into the second plate.

A strongly low frequency case is shown in Figure \ref{Fig09}, where $\beta=0.5119\,\mbox{m}^{-1}$ ($\omega=0.0408\,\mbox{rad/s}$). 
In part (a) of the figure it is shown that the wave does not propagate in the rectangular plate and the incident wave $W_I$ is almost entirely reflected into $W_R$. In part (b) of Figure \ref{Fig09}, we add a system of 7 snail resonators in the vicinity of the ligament, as shown in the inset of Figure \ref{Fig08}a. The geometry of the resonators is described by Eq. (\ref{eqn003}), with $R_0=0.35\,\mbox{m}$ and $R_1=0.7\,\mbox{m}$. Such a system of resonators, when activated by the incident wave $W_I$ is capable to excite vibrations in the finite rectangular wave enhancing the transmission.  
In parts (c) and (d) of Figure \ref{Fig09} the same experiment is repeated at the higher frequency $\beta=1.4547\,\mbox{m}^{-1}$ and a similar result is obtained, namely negligible vibrations for the rectangular plate for the homogeneous case and enhanced vibrations for the case with the addition of resonators. 

We stress the fact that $\beta=1.4547\,\mbox{m}^{-1}$ is very close to the first eigenfrequency of the snail resonators, namely $\beta=1.4469\,\mbox{m}^{-1}$, while $\beta=0.5119\,\mbox{m}^{-1}$  is different not only from any eigenfrequency of resonators, but also from the eigenfrequencies of the mechanical system composed by the rectangular plate plus the tiny ligament with or without perforations and resonators.

\subsection{Vibration suppression in a waveguide}

\begin{figure}[ht!]
\centering
{\includegraphics[width=12cm,angle=0]{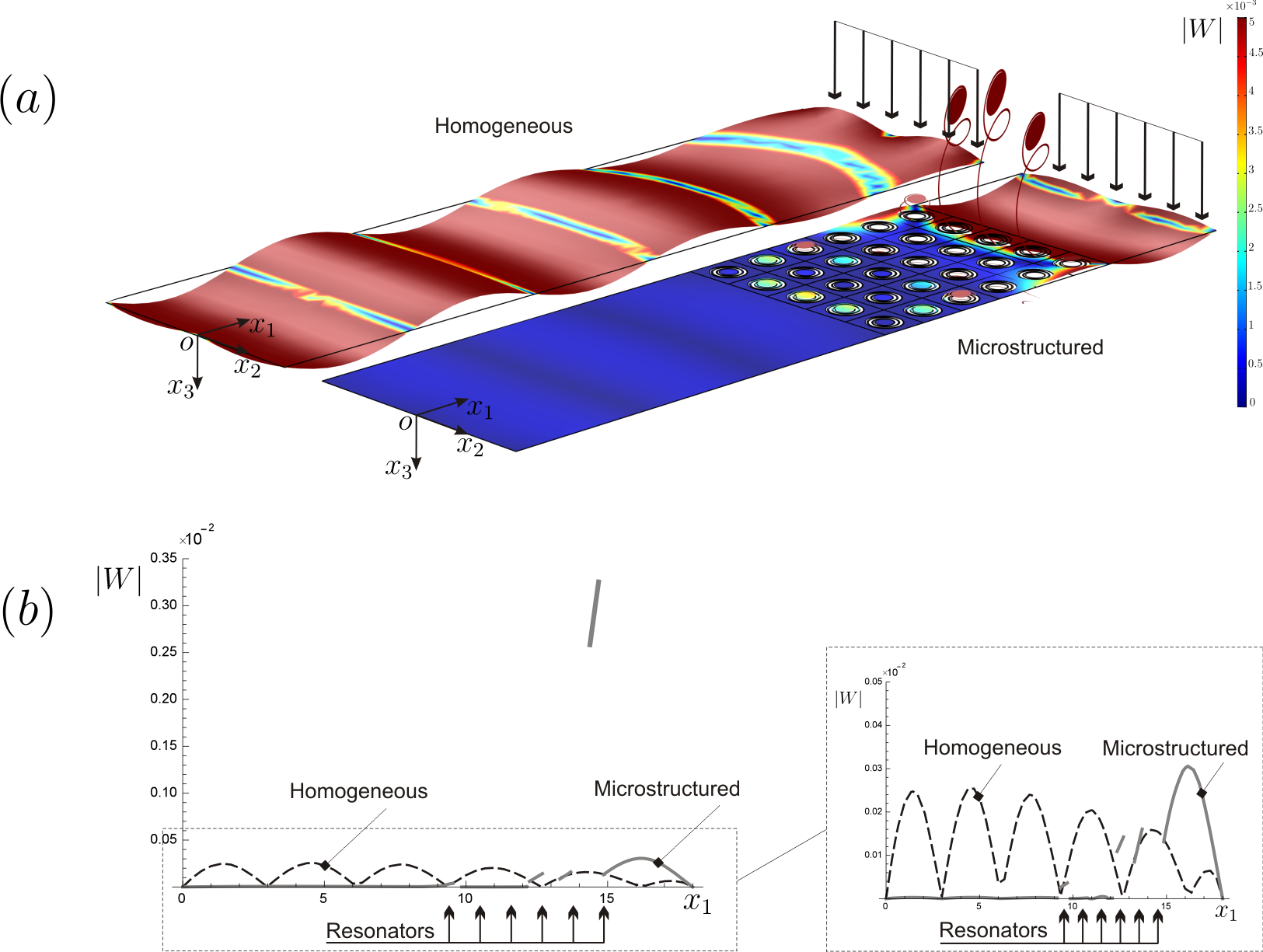}}
 \caption{\footnotesize (a) Vibrations of a homogeneous plate vs homogenous plate with microstructured interface composed of $6\times 5$ resonators. (b) Transverse displacement $|W|$ along the axis $x_2=0$. Homogenous plate is given in black dashed line, plate with microstructured interface in continuous grey line. Displacement in the resonator inclusions are evident.}
   \label{Fig12}
\end{figure}

\begin{figure}[ht!]
\centering
{\includegraphics[width=12cm,angle=0]{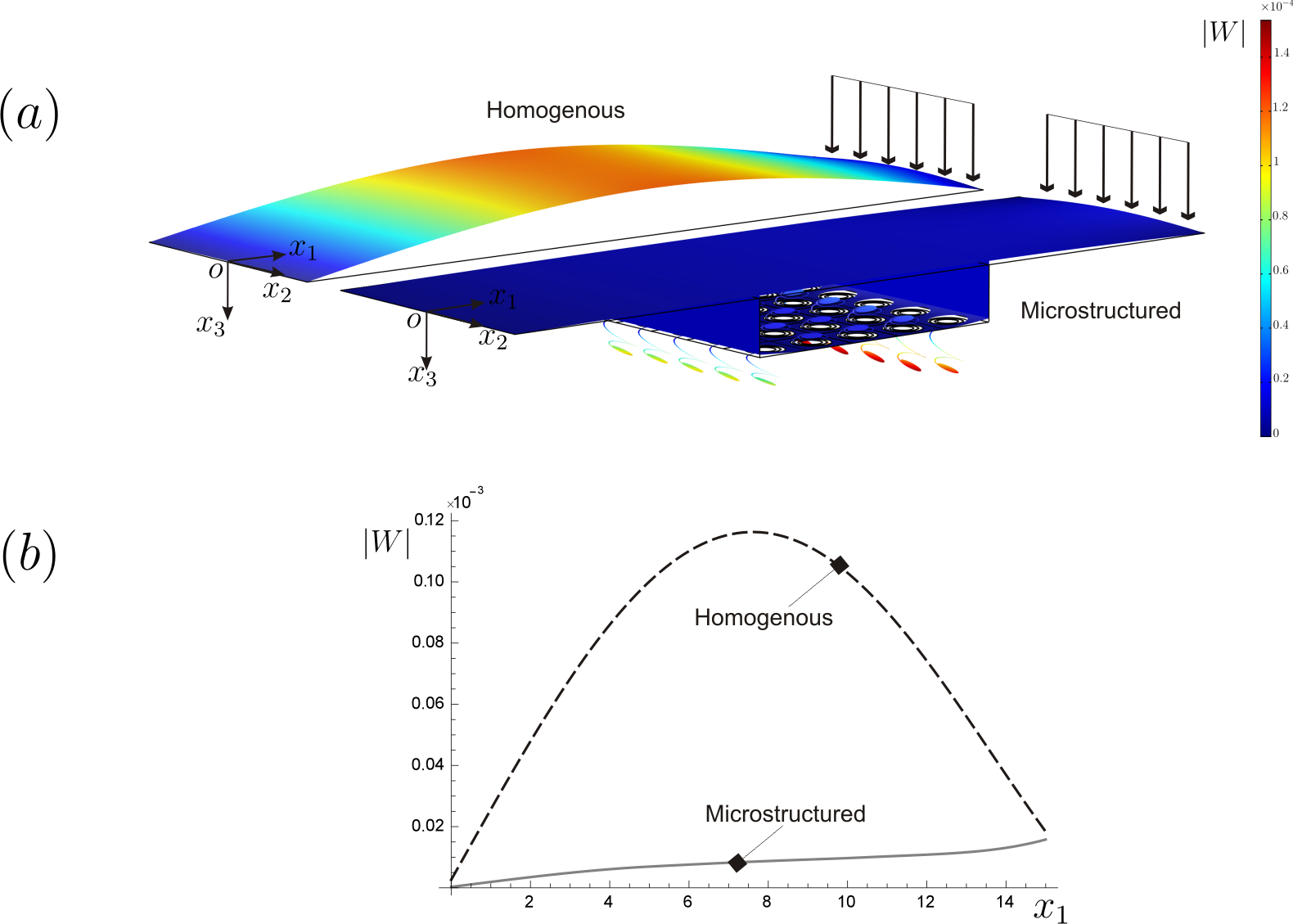}}
 \caption{\footnotesize (a) Vibrations of a homogeneous plate vs homogenous plate with microstructured by-pass system composed of $5\times 5$ resonators. (b) Transverse displacement $|W|$ along the axis $x_2=x_3=0$. Homogenous plate is given in black dashed line, plate with by-pass system in continuous grey line.}
   \label{Fig13}
\end{figure}

As a final example of possible applications of the microstructured medium, we propose the design of a lightweight �wave bypass� structure which is capable to divert large amplitudes vibrations away from load-bearing elements. 

We start with a more standard approach in Figure \ref{Fig12},  which involves the study of a finite structure with repetitive units as a perfect periodic structure.

The steel plate has dimensions $18\,\mbox{m}\times 5\,\mbox{m}$, thickness $h=1\,\mbox{mm}$ and it is simply supported at $x_2=\pm2.5\,\mbox{m}$, $x_1=0,\,3,\,6,\,9,\,12,\,15,\,18\,\mbox{m}$ and $x_2=0\,\mbox{m}$, $x_1=0,\,18\,\mbox{m}$. The structure is subjected to an harmonic transverse edge load at $x_1=18\,\mbox{m}$ having magnitude equal to $1\,\mbox{N}/\mbox{m}$. In Figure \ref{Fig12} the load is vibrating with frequency $f=0.25\,\mbox{Hz}$, i.e. 
$\beta=0.3507\,\mbox{m}^{-1}$, which corresponds to the first eigenfrequency of the finite plate or to the frequency of the first stationary mode within the dispersion diagram of the periodic homogeneous  plate composed by a repetitive unit of dimension $3\,\mbox{m}\times 5\,\mbox{m}$. The vibration mode is presented on the upper part of Figure \ref{Fig12}a.

In order to reduce vibration on the left part of the plate we introduce an interface composed of $6$ lines of $5$ square units having the geometry shown in Figure \ref{Fig01}a. The microstructured plate is visible on the lower part of Figure \ref{Fig12}b.  
The geometrical parameters of the resonators are predesigned following the asymptotic model reported in Section \ref{SemiAnaly} in order to match with the first eigenfrequency, associated with a translational vibration of the inclusion, the targeted frequency $\beta=0.3507\,\mbox{m}^{-1}$. 
In this step we consider a single isolated snail resonator. 

A second fine tuning of the geometrical parameters leads to full coupling of the resonators with the plate. This is done analyzing numerically the full structure with the microstructured interface.
In particular, we have chosen to change the in-plane thickness of the ligament in the resonators in order to obtain the desired wave filtering. 
The final geometrical properties are: $R_0=0.175\,\mbox{m}$, $R_1=0.35\,\mbox{m}$, in plane thickness $s=21.875\,\mbox{mm}$. 

Such a system of resonators is capable to open a tiny band-gap at $\beta=0.3507\,\mbox{m}^{-1}$. The vibration amplitudes shown in Figure \ref{Fig12} demonstrate the capability of the interface to block the wave propagation within the plate. The incoming waves are reflected by the interface and the displacement decay exponentially fast. Such a system is highly effective and does not require a heavy variation of the original structure. In the proposed case the final structure is even lighter than the original one. 

The drawback of the proposed approach is that the energy is reflected back by the interface and still excite the region $x_1\ge 15\,\mbox{m}$ ahead the interface. The displacements in the inset of part (b) of Figure \ref{Fig12} show that the amplitude of vibration in this region is larger than the homogeneous case.   

In Figure \ref{Fig13} we propose an alternative approach which is capable to re-route wave propagation within the whole structure.
In particular, we consider an initial steel rectangular plate having dimensions $15\,\mbox{m}\times 5\,\mbox{m}$, thickness $h=25\,\mbox{mm}$ and simple supports at $x_1=0,\,15\,\mbox{m}$ and $x_2=\pm 2.5\,\mbox{m}$.

The microstructured plate is now connected in `parallel' to the main structure in the central region $5\,\mbox{m}\,\leq x_1\,\leq 10\,\mbox{m}$. 
The design procedure of the microstructure plate follow the same scheme detailed above, with a first predesign step on a single resonator followed by the analysis of the full structure shown in the bottom part of Figure \ref{Fig13}a in order to obtain full coupling between the initial plate and the attached by-pass system.

The comparisons  between the deformed shapes shown in part (a) of Figure \ref{Fig13} and displacement magnitudes along the axis $x_2=0$ shown in part (b), reveal the drastic reduction of the displacement amplitudes. We stress that the amplitude reduction is extended to the entire initial structure, which has changed the vibration mode under the excitation of the edge load at $x_1= 15 \,\mbox{m}$. In the modified structure the upper plate displays low amplitude vibrations and wave propagation is forced to be redirected into the system of resonators. 

Finally, we stress that the study is within the elastic range and we do not consider any energy dissipation effect. Clearly, the energy stored in the resonators could be stored or dissipated. It is also evident that damping system can be efficiently placed around the snail-resonator positions.

\section{Conclusions}

A newly designed \emph{platonic} crystal is proposed. 
The microstructured medium can lead to wave localization, wave trapping and edge waves. It has also been applied to re-route waves in order to produce low amplitude vibration on the main structure.

The design of the model is simplified by the possibility to estimate analytically or numerically resonance frequencies of the inertial resonators. Simple geometrical parameters may be used to have specific effects at targeted frequencies. 

The proposed structured plate is very attractive for technologically applications since it is a single phase material and it can be produced at low cost by existing technologies. Standard techniques are additive manufacturing at small scale and water jet cutting or laser cutting on homogeneous plates at larger scales.

\clearpage
\setcounter{equation}{0}
\renewcommand{\theequation}{{A}.\arabic{equation}}
\begin{center}
{\bf Appendix A: modified resonator}\\
\end{center}

\begin{figure}[!ht]
  \centerline{
  \includegraphics[width=9cm]{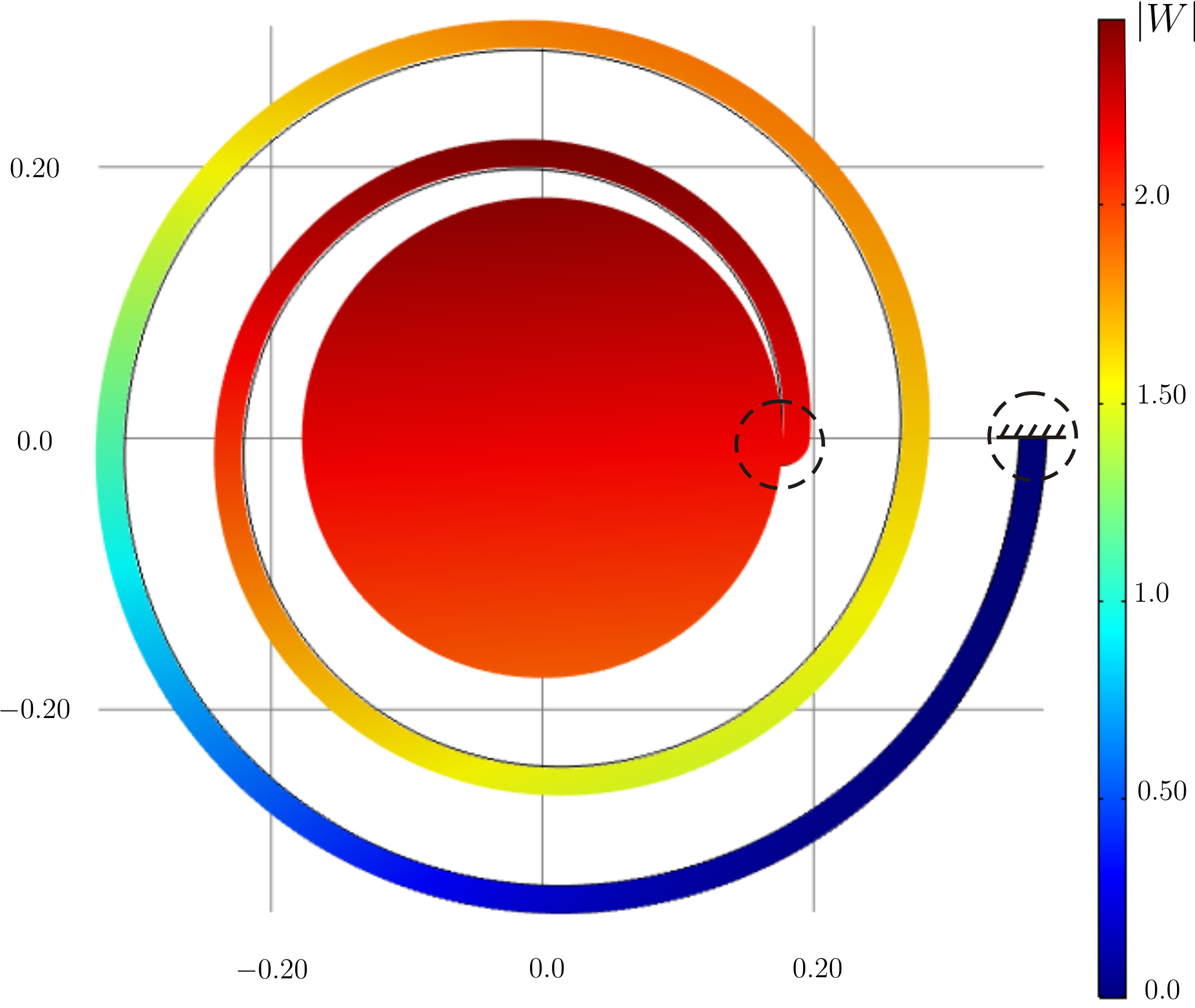}}
  \caption{\footnotesize Modified geometry of the resonator. Spiral ligament ends have been modified. The first eigenmode is shown.}
  \label{FigApp001}
\end{figure}

The modified resonator shown in Figure \ref{FigApp001} has also been implemented in \emph{Comsol Multiphysics}\textsuperscript{\textregistered}. It has different connections at the ligament ends which better mimic the constraints of the monodimensional analytical model. The corresponding eigenfrequncy is $\beta_1=0.9331$ m$^{-1}$, which is in excellent agreement with the analytical prediction shown in Table \ref{Table01}. 

\end{document}